\pdfoutput=1

\documentclass{sig-alternate-05-2015}

\usepackage{balance}  
\usepackage{booktabs}
\usepackage{multirow}
\usepackage{graphics}
\usepackage{graphicx}

\usepackage{color}
\newcommand{\RED}{}

\begin{document}






%

\title{Visual TASK: A Collaborative Cognitive Aid for \\Acute Care Resuscitation}
%
%
%
%
%

\numberofauthors{2} 
%
\author{
%
%
Michael J. Gonzales$^{1}$, Joshua M. Henry, MD$^{2}$, Aaron W. Calhoun, MD$^{2}$, and Laurel D. Riek, PhD$^{1}$\\
\and
\alignauthor
       \affaddr{$^{1}$Dept. of Computer Science and Engineering}\\
       \affaddr{University of Notre Dame}\\
       \affaddr{Notre Dame, IN 46556}\\
       \email{\{mgonza14, lriek\}@nd.edu}\\
\alignauthor
       \affaddr{$^{2}$Dept. of Pediatrics, Division of Critical Care}\\
       \affaddr{University of Louisville}\\
       \affaddr{Louisville, KY 40202}\\
       \email{\{joshua.henry, aaron.calhoun\}@louisville.edu}\\
 }

\maketitle
\begin{abstract}
Preventable medical errors are a severe problem in healthcare, causing over 400,000 deaths per year in the US in hospitals alone. 
In acute care, the branch of medicine encompassing the emergency department (ED) and intensive care units (ICU), error rates may be higher to due low situational awareness among clinicians performing resuscitation on patients. 
To support cognition, novice team leaders may rely on reference guides to direct and anticipate future steps. However, guides often act as a fixation point, diverting the leader's attention away from the team. 
To address this issue, we conducted a qualitative study that evaluates a collaborative cognitive aid co-designed with clinicians called Visual TASK. 
Our study explored the use of Visual TASK in three simulations employing a projected shared display with two different interaction modalities: the Microsoft Kinect and a touchscreen.
Our results suggest that tools like the Kinect, while useful in other areas of acute care like the OR, are unsuitable for use in high-stress situations like resuscitation.
We also observed that fixation may not be constrained to reference guides alone, and may extend to other objects in the room.
We present our findings, and a discussion regarding future avenues in which collaborative cognitive aids may help in improving situational awareness in resuscitation.
\end{abstract}

\category{H.5.m.}{Information Interfaces and Presentation
  (e.g. HCI)}{Miscellaneous}{}{}
\vfill
\section{Introduction}
As many as 400,000 people die each year due to preventable medical errors in the United States in hospitals alone \cite{James2013}.
This makes preventable medical errors the third leading cause of death, and equates to two-thirds of those who die every year from heart disease and all form of cancer \cite{CDC}.
What is most troubling about this fact is that most of these errors, if not all, can be avoided.

Preventable medical errors can be segmented into one of five categories.
Errors of commission are mistakes that harm a patient. Errors of omission are errors that result from forgetting to perform an action. Errors of communication are errors that are caused by miscommunication between team members. Contextual errors are those that occur when a patient's condition affects the course of treatment. Finally, diagnostic errors occur when ailments affecting a patient are misdiagnosed \cite{James2013}.

Many of these errors may be attributed to fatigue, a lack of knowledge due to training, temporary workers, and issues with patient hand-offs. However, a majority of others may be occur due to a lack of situational awareness (SA) \cite{Baker2006, Fioratou2010}.
SA is the ability to perceive elements in the environment and make use of that information to inform future actions \cite{Endsley1995, Endsley2000}.
Errors of commission, omission and communication may result from a lack of SA, particularly when clinicians fixate on one specific aspect or event of a situation.
According to Fioratou et al. \cite{Fioratou2010a}, fixation errors occur when practitioners  ``concentrate solely upon a single aspect of a case to the detriment of other more relevant aspects.''
A reduction of fixation errors may help improve SA in dynamic environments like acute care \cite{Fioratou2010a, Xiao1995}.

In addition to SA, communication and coordination are critical in acute care, because it involves teams of inter-professional clinicians working together to address a patient's condition \cite{Risser1999}.
Inter-professional teams are defined as a group of clinicians from different disciplines and areas of expertise (e.g., hospitalists, emergency physicians, intensivists, nurses).
Acute care is a challenging environment to maintain SA due to frequent changes in the patient's condition, multiple information exchanges between team members, and a high-cognitive load due to efforts to anticipate future steps.

One of the most common procedures conducted in acute care is resuscitation, or a code, which is the act of insuring that the body is receiving the various elements it needs in order to operate normally.
Codes are highly time-sensitive, with clinicians entering the scene at different points in time and taking on varied roles in an effort to provide support to the team. Roles can rapidly change over time, leading to decreased SA due to interrupted flows of information \cite{Flin2004}.

Codes may rapidly progress to life threatening situations that are difficult to manage.
Depending on the environment in which a code occurs (for example, a general ward as opposed to the ICU), as many as 20 individuals may end up entering a scene, complicating the situation rather than helping \cite{Gonzales2015}.
Previous work in the area of acute care resuscitation has shown that fixation may be a common challenge for novice clinicians \cite{Gonzales2015,Nelson2008}.
Fixation errors are areas of hyperfocus on one aspect in a situation to the exclusion of other necessary information \cite{Gosbee2010}.
In resuscitation, fixation occurs often due to an overreliance on reference guides by novices, leading to delays in treatment, including CPR and drug administration, or  incorrect diagnoses \cite{Nelson2008}.

\begin{figure}[t]
\centering
\includegraphics[width=.85\columnwidth]{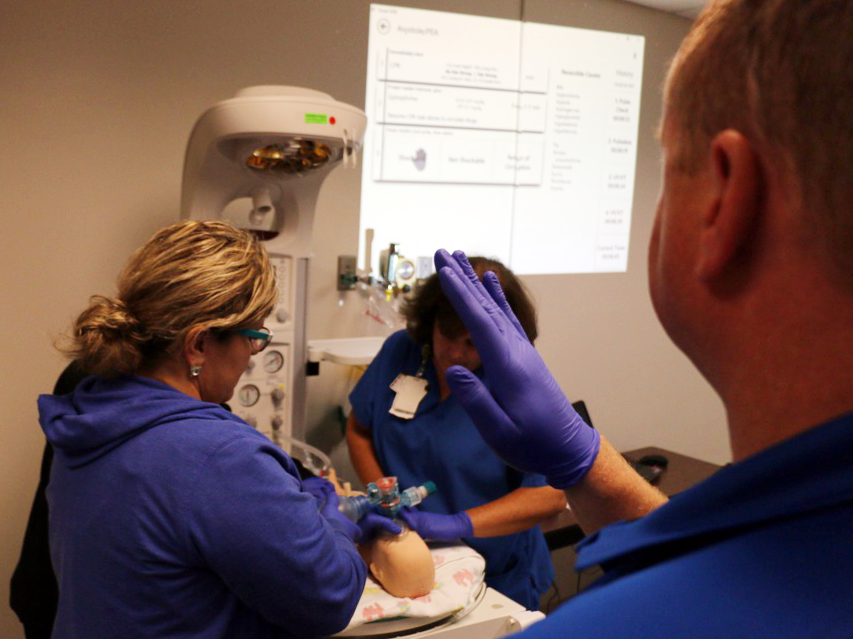}
\caption{An example of our collaborative cognitive aid used with the Kinect in a code simulation.}
\label{fig:fig1}
\end{figure}

In an effort to address these challenges, we introduce the Visual TASK (Team Awareness and Shared Knowledge) system.
The aim of Visual TASK is to reduce fixation errors in resuscitation by providing a projected shared display, which highlights key information about successive steps during resuscitation.
The tool is composed of a touch-screen laptop computer with a projected display, and includes a Microsoft Kinect as an alternative form of interaction for training purposes (see Figure \ref{fig:fig1}).
Visual TASK was co-designed with clinicians from three different US health institutions.

In this paper, we present a qualitative study that evaluates Visual TASK with interprofessional clinicians at a large midwestern children's hospital, within \textit{in-situ} simulated codes. We explored the following research questions:
1) How does a collaborative cognitive aid affect fixation in simulated codes?
2) How do clinical learners perceive the tool, and how does a collaborative cognitive aid help promoting a common understanding of a code situation?

Our findings suggest that there may be a number of avenues in which collaborative cognitive aids can aid in improving fixation, provided the right tools are used in the right environment.
For example, we found that tools like the Microsoft Kinect, which have been explored in the context of the operating theatre \cite{o2014}, may not be ready for higher-paced collaborative environments.
In addition, we observed instances in which novice clinicians focused vital time on seeking answers from tools in the room, rather than thinking critically about the situation facing them.
We also found that by providing only information relevant to the current situation, shared cognitive aids like Visual TASK may be helpful in providing the team a cohesive mental model of a situation.
Overall, these findings point to the potential advantages of shared resource displays in resuscitation.

\section{Background}
\subsection{Acute Care Resuscitation Overview} 
Acute care resuscitation is the act of returning the patient's physiology to a state that can fully support the needs of the patient's body. 
This process requires specially trained physicians, such as emergency medicine or intensive care physicians.
Resuscitation requires a large team of health care providers that each focus on a specific job, such as giving medications or performing cardiopulmonary resuscitation (CPR).
Team members may enter a code scene at any time, and team sizes can range from four to twenty \cite{Gonzales2015}.

In order for a code to run effectively, a team leader directs the code, and must maintain SA of the room, identify ongoing management strategies likely to be successful, and direct the team's tasks \cite{ACLS, PALS, AHRQ}.
This role requires the ability to manage multiple streams of information simultaneously in order to work in an efficient and meaningful way.

Each team member is assigned one job in the code to ensure it is adequately addressed and not compromised by other ongoing tasks.
Common team roles include: managing the airway, performing CPR, preparing and giving emergency medications, monitoring and evaluating the patient's vital signs, and documenting tasks.

Due to the complexity of resuscitation, trainees typically follow standardized, evidence-based algorithms published by the American Heart Association (AHA).
Two such algorithms maintained by the AHA include Pediatric Advanced Life Support (PALS) \cite{PALS} for children, and Advanced Cardiac Life Support (ACLS) for adults \cite{ACLS}.

Clinicians typically take a course to be certified in ACLS and PALS. PALS reference guides, also referred to as simply ``PALS cards,'' are given to clinicians as a part of the course for use in both simulated and real-world resuscitation (See Figure \ref{fig:fig2}, left).
These cards act as cognitive aids for clinicians to understand the reasons for a code, along with algorithms for different patient conditions that vary based on pulse, rhythm, and other vitals.
While helpful in codes, PALS cards are only useful if they are interpreted and executed correctly by the team leader.
Oftentimes, clinicians make mistakes in applying the appropriate algorithm, leading to diagnosis errors and errors of commission.

\subsection{Situational Awareness and Fixation}
Research in healthcare has focused on the ability for clinical teams to maintain SA and common mental models in high stress environments.
This can lead to cognitive lockups, which are failures due to missing important information due to unttended informational resources.

For example, low-tech tools like Situation Background Assessment Recommendation (SBAR) are designed so that teams of clinicians can maintain common mental models and communicate effectively \cite{Haig2006}.
Researchers have also developed a wide breadth of collaborative low-tech and high-tech  tools aimed at improving SA, including the use of whiteboards \cite{Xiao2001}, mobile tools for hand-offs \cite{Bossen2008}, and collaborative cognitive aids for trauma environments \cite{Sarcevic2012}.
Despite their potential, many of these tools are built for low intensity situations or roles, and can themselves become fixation points.

\begin{figure}[t]
\centering
\includegraphics[width=.85\columnwidth]{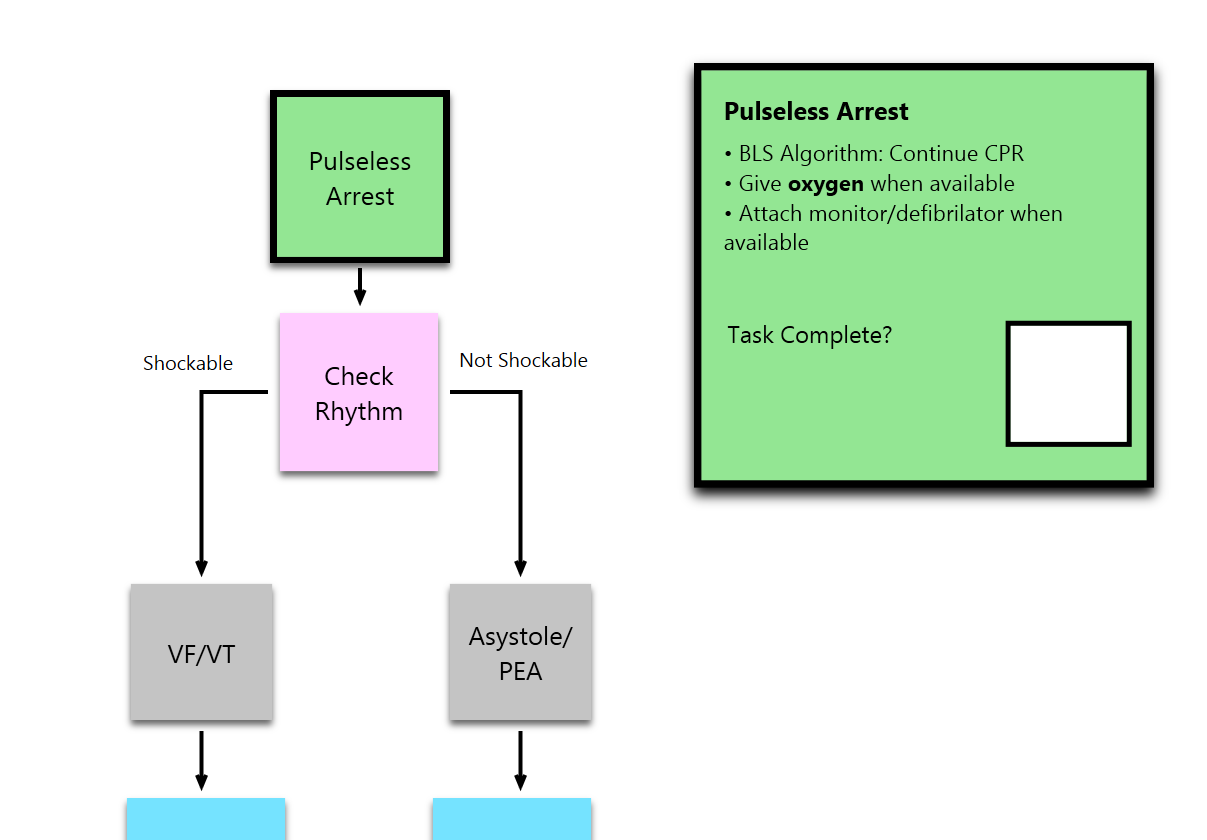}
\caption{Our original prototype for Visual TASK that mimicked the PALS design.}
\label{fig:figoldvt}
\end{figure}

Low fidelity tools, such as checklists are used commonly in operating rooms for clinicians in surgeries to maintain SA \cite{Healey2010, Gawande2010}.
They are often used as a way for clinicians to verify that tasks are completed thoroughly.
However, checklists and other reference aids may be difficult to use in resuscitation.
Previous work showed that fixation errors and cognitive lockups occurred commonly in dynamic environments like resuscitation \cite{Xiao1995, Sarcevic2012}.

{\RED{
The literature has also explored the use of collaborative technology in crisis management \cite{Wu2014, Cirimele2014}. This work showed that using a step-wise approach to tasks significantly reduced errors in individual assessments with clinicians, and that layout can have a significant effect with cognitively processing cognitive aids. However, one gap still remains  with evaluating technology like cognitive aids in collaborative simulation.}}

Instead, dynamic situations may require solutions that can accommodate the changing and varied situations clinicians face.
Codes are one such case due to the simultaneous processes that must occur in unison and the rapidly changing condition of the patient.
This led us to the design of Visual TASK with an approach that presents only the relevant tasks of a situation on a shared display for teams.

\section{Visual TASK: System Design}
Previously, we conducted an ethnographic field study that involved environmental walkthroughs, observations of mock codes, and contextual inquiries at two regional and one urban US hospital \cite{Gonzales2015}.

This work aided in helping us to identify a number of problems clinicians encounter during resuscitation, both in simulation and the real-world.

\begin{figure*}[t]
\centering
\includegraphics[width=.95\textwidth]{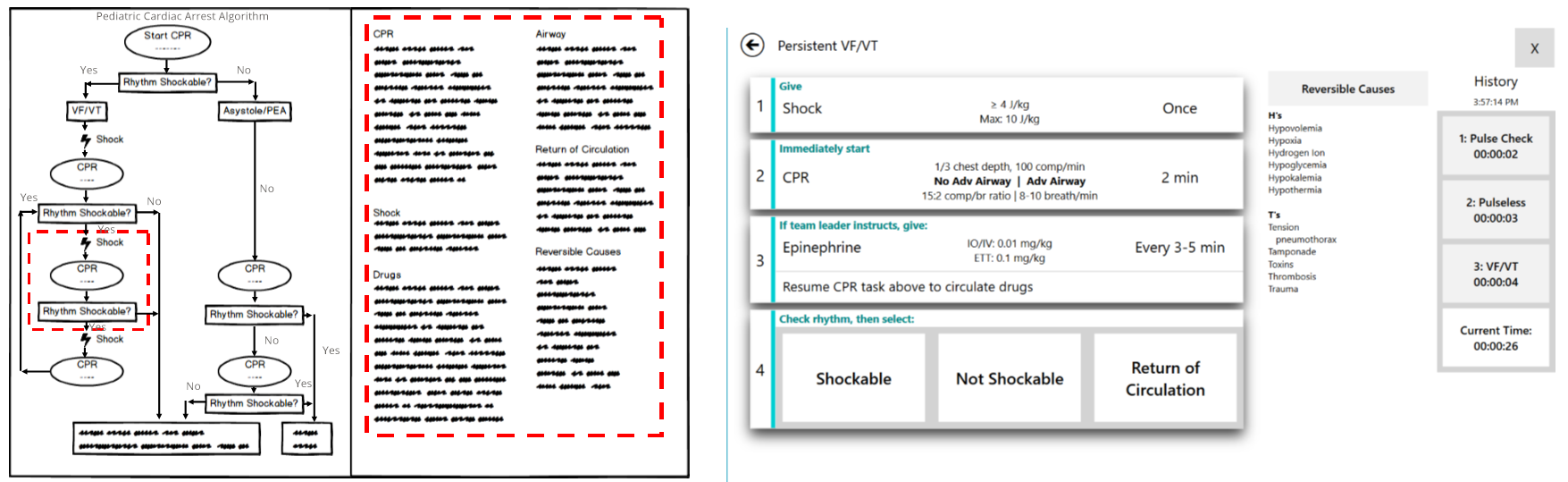}
\caption{Left: Part of a PALS reference guide for the Pulseless Arrest Algorithm. Right: Visual TASK, our redesigned PALS guide. The red/dashed highlighted region shows the current split of information on PALS cards, whereas our redesign pairs this information with its associated task.}
\label{fig:fig2}
\end{figure*}

Some of these challenges included an over-reliance on reference guides by novices, leading to cognitive lockups and fixation errors that also disrupt the team.
Clinicians also had problems communicating when new entrants came to the scene, causing frequent repetition of the same information for each individual.
Documentation was another issue, with current electronic health record and paper-based solutions failing during codes, causing clinicians to rely on paper towels to record information.

Following the study, we decided to explore potential solutions to the problems clinicians faced.
Based on input from clinical collaborators, we focused on the design of a new type of cognitive aid that could potentially aid in reducing fixation errors and help new entrants as they enter a code scene.
This also provided us with a foundation upon which we could potentially augment documentation at a later point.

In order to design this new cognitive aid, we engaged in a longitudinal, iterative design process with clinicians that engaged in our ethnographic work.
Our design process included nine interprofessional clinicians, across three different US hospitals.
Clinicians provided sketches, drawings, clinician-developed mock-ups, and provided important considerations for each role in codes through semi-structured interviews.
This process also involved an analysis of the environments, artifacts, and common problems observed by clinicians in both simulated and real-life code situations.

A large part of our design process involved redesigning the PALS reference guide and the way it presented information.
Our original design closely mimicked the PALS card structure, individually going through tasks in a flowchart manner (see Figure \ref{fig:figoldvt}).
This eventually changed to a step-wise grouping of tasks that occur simultaneously, so that each clinician in the room can gather information about their current task (see Figure \ref{fig:fig2}).
This includes timing/duration and amount/dosage for each task happening between heart rhythm checks.
Throughout the design process, we developed a total of 20 different designs aimed at refining the current algorithms used in PALS Pulseless Arrest situations \cite{PALS}.
Following wireframe and mockup generation, we conducted cognitive walkthroughs with clinicians, providing a scenario similar to what clinicians might experience in actual code situations.

Algorithms are selected by answering important questions related to the patient's condition.
The very first question clinicians see is: Does the patient have a pulse?
The algorithm then progresses to a follow-up screen with other tasks/question as dictated by AHA's guidelines \cite{ACLS, PALS}, eventually presenting tasks associated with the appropriate algorithm for the patient's condition.

In addition to redesigning the algorithms, selecting suitable hardware was another major design consideration, especially since each hospital varied in space and resources \cite{Gonzales2013b}.
Because communication and information exchanges are so vital in resuscitation, we focused on a shared display solution that could provide a cohesive data display for the various tasks performed by teams during codes.
We used a projected display to reduce the probability of introducing new hazards, such as the possibility of knocking over new monitors or drawing wires across the room.
This solution can be used portably, on multiple surfaces, and in different contexts.

Due to clinical educators' desire to train clinicians to keep their peripheral vision available while referencing information, we implemented the system with two different interaction modalities: The Microsoft Kinect v. 2, and a touch screen.
The Kinect was intended as a method for clinicians to gesturally interact with the system, thus reducing the possibility of a division of attention between multiple screens.
We decided to utilize the Kinect because it has been employed in other acute care environments, such as the operating room \cite{o2014, Ruppert2012}.

By involving multiple institutions in our design process, we realized that there were environments in which a Kinect might not be feasible due to space constraints.
Some institutions were environmentally constrained, with only 100 sq ft. per patient bed for a code team to perform their tasks.
This spatial constraint significantly impacts the capability for clinicians to interact gesturally with the Kinect.
For these cases, we also utilized touch-screen display as an alternate interaction modality.

Once we finalized a design that satisfied our clinical collaborator's constraints, we proceeded to conduct a qualitative field trial of Visual TASK at Midwestern Children's Hospital.

\section{Evaluation Methodology}

\subsection{Study Setting}
\begin{figure*}[t]
\centering
\includegraphics[width=.8\textwidth]{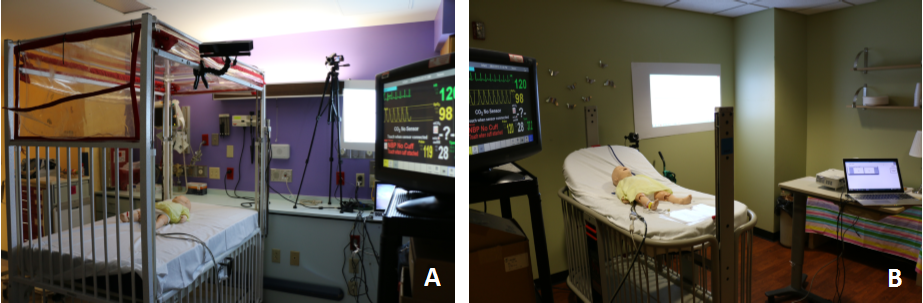}
\caption{We conducted three \textit{in situ} resuscitation simulation sessions in the hospital's Pediatric Intensive Care Unit (PICU). A) The room layout for Sessions 1 and 2. B) The room layout for Session 3.}
\label{fig:fig3}
\end{figure*}

We conducted a total of three simulation sessions aimed at exploring the use of Visual TASK in acute care resuscitation.
The study was conducted in the pediatric ICU of the children's hospital, see Figure \ref{fig:fig3}.

\subsection{Participants} 
We explored Visual TASK with a total of 23 clinicians (Average age: 31.55, SD: 11.42, 19 female, four male).
Clinicians from multiple professions participated in our study, including 5 physicians, 11 nurses, 5 respiratory therapists, and one medical student. 
With the exception of the medical student, all had participated in a PALS simulation at least once before and have worked in the healthcare industry between 3 months, to 43 years.
Each session had between seven and eight participants.

The hospital's program requires clinicians to engage in mock codes at different levels of frequency per year depending on the clincian's unit and profession.
Each session was already pre-scheduled as part of the hospital's standard simulation program, and we did not in any way modify participant composition or the educational goals of the program to explore the use of the tool.

Participants provided informed consent, in accordance with the Institutional Review Board (IRB) at the University of Louisville.
Participants who did not wish to participate were given the opportunity to reschedule their exercise, or conduct a simulation as a part of standard practice.

\subsection{Simulation Structure}
For the three sessions, we instructed all team leaders to use the system only if they needed a reference guide for the code they managed.
All sessions followed standard simulation procedures of the hospital program, which included: an introduction and tutorial, the simulation sessions themselves, and a debriefing discussing the team's performance.

Each session followed one of two  simulation types.
\textit{Rapid recognition simulations} involve each person rotating into the team leader role.
Each of these simulations lasts approximately five minutes, and involves a new team leader recognizing the patient condition and organizing the team accordingly.
\textit{Standard simulations} involve one prolonged code that lasts approximately 15 minutes, where one person is the team leader for the duration.

Prior to each session, we provided a five minute tutorial at the start of the simulation to help clinicians learn how to use the tool. All sessions used a projected display as the cognitive aid for sharing the team's progress as opposed to standard PALS cards.

For sessions that employed the Kinect, team leaders received a short tutorial of the gestures used to control the device, including gaining control by raising their hand over their head, and selecting items by pushing their hand forward. We allowed team leaders to practice using the tool for five minutes prior to the start of the simulations.

Due to scheduling and educational constraints at the hospital, it was not possible to strictly control the type of simulation conducted per session (i.e., rapid or standard), nor to conduct a fourth session using the touchscreen in a standard simulation. However, we decided the overall ecological validity of our study was far more important than controlling for these factors, and we report our results accordingly. Thus, Sessions 1 and 2 employed the Kinect, and Session 3 a touch screen. Sessions 1 and 3 employed a rapid recognition structure, and Session 2 a standard one. 

After each of the three sessions, participants engaged in the standardized post-simulation debriefing process used at the hospital.
Following this, clinicians responded to a qualitative survey regarding their perceptions of Visual TASK and its effect on team dynamics.

\begin{table}[t]
\centering
\caption{Annotation categories. Annotators primarily focused on time spent focused on different aspects of the code environment. }
\label{annotations}
\resizebox{.85\columnwidth}{!}{%
\begin{tabular}{ll}
\\ \hline
\multicolumn{2}{c}{Annotations}                                                                                                                                                                        \\ \hline
                                                        
\begin{tabular}[c]{@{}l@{}}Interaction/\\ Recognition Challenges\end{tabular} & \begin{tabular}[c]{@{}l@{}}\\Directed glances focused \\ on the shared display.\\\end{tabular}                                      \\ 
Shared Display Focus                                                          & \begin{tabular}[c]{@{}l@{}}\\Focus and challenges with \\ interaction and gaining \\ recognition from the system.\\\end{tabular}    \\ 
Monitor References                                                            & \begin{tabular}[c]{@{}l@{}}\\Focus of attention on the \\ patient monitor for vitals.\\\end{tabular}                                \\ 
Team and Patient Focus                                                        & \begin{tabular}[c]{@{}l@{}}\\Focus/monitoring of team \\ members and the patient.\\\end{tabular}                                    \\ 
Other Attentional Focus                                                       & \begin{tabular}[c]{@{}l@{}}\\Other attentional distractions \\ away from the team, \\ patient, or an ongoing action.\\\end{tabular} \\ 
\end{tabular}%
}
\end{table}

\subsection{Video Collection and Analysis}
In addition to the written questionnaire, all sessions were video-recorded to enable detailed analysis of how participants responded to and used the system.
Following all sessions, the research team, composed of computer science experts and PALS experts conducted informal video walkthroughs to discuss the tool's affect on team dynamics, input into the types of labels annotators would focus on during analysis, as well as educator perceptions.

Two independent annotators analyzed video data based on the themes described in Table \ref{annotations} using deductive coding steps, and had high inter-rater reliability (\textit{k-alpha = 0.95}).
Annotators labeled all attentional focus for the team leader, and coded references to the system display for other members of the team greater than 500ms using video annotation tools.
According to Wickens et al. eye fixation typically occurs in glances over 500ms \cite{Wickens1992}.

Glances for each object (the shared display, patient monitor, and team/patient) consisted of visual eye movement (action units 61, 62, 63, and 64) and head movement (action units 51, 52, 53, and 54) in the direction of the object, until the point of focus shifted away.
Other attentional glances were those directed at other objects in the scene that were not a part of the code during an ongoing task.

\section{Key Findings}

\subsection{Video Observations}
\begin{figure*}[t]
\centering
\includegraphics[width=.95\textwidth]{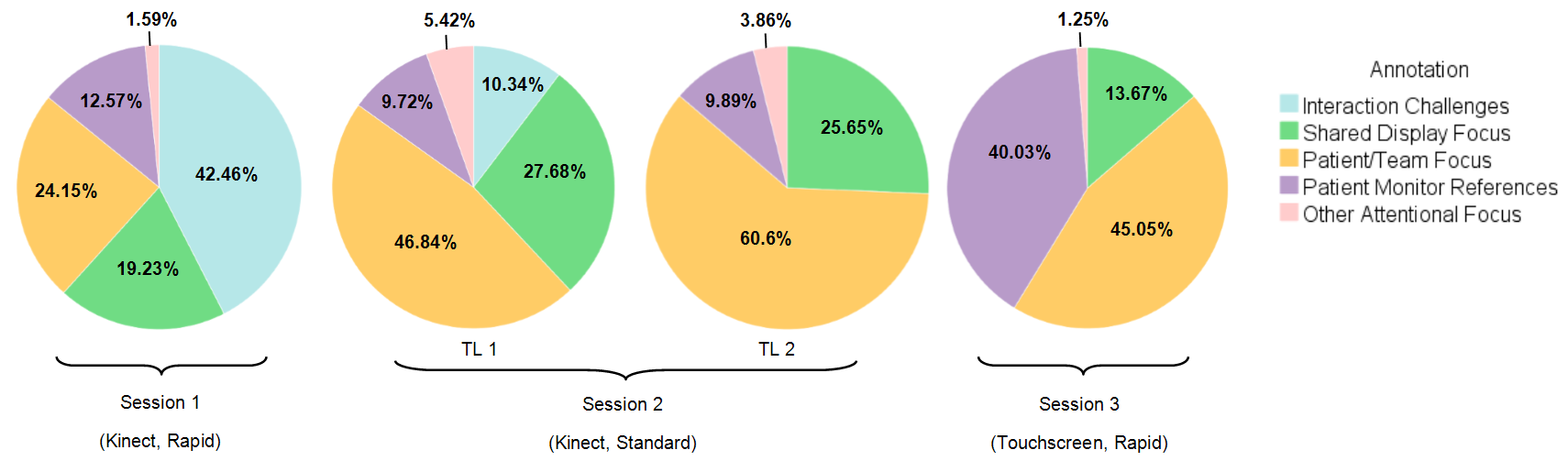}
\caption{The division of attentional focus for team leaders in each session. Session 2 contained two simultaneous team leaders. Division of attention is more important than total time here, since simulation times may vary in length due to instruction and debriefing.}
\label{fig:graph}
\end{figure*}

Figure \ref{fig:graph} shows the distribution of time spent by clinicians among different attentional foci.
The following reported data is based on estimates of attentional focus time as measured by the two independent annotators. While these annotations are not perfect, they represent a good estimate of attentional focus, and are similar to a method used by Bach et al. to estimate attentional focus and fixation \cite{Bach2008}. 

During Session 1, team leaders focused the majority of their time (approximately 42.46\%) getting the Kinect to recognize them when using Visual TASK.
Team leaders spent the rest of their time attending to the team and patient (24.15\%), followed by references to the shared display (19.23\%) and patient monitor for vitals (12.57\%).

During Session 2, two clinicians ended up acting as a pair of team leaders (which sometimes occurs during real-world scenarios, too).
Having multiple team leaders can lead to confusion for the rest of the team due to contradictions in direction and miscommunication.
Instructors let these scenarios play out when they manifest in simulation to highlight and discuss potential problems clinicians encounter.

In Session 2, one team leader ended up controlling the system, while the other focused more on directing the team verbally.
The team leader controlling the device spent the majority of their time focusing their attention on the team (approximately 47\%), approximately 27.68\% of their attention on the shared display, and 10.34\% focused on interaction. The remainder of their time focused on the patient monitor (9.72\%) and other attentional glances (5.42\%).

The second team leader spent the most time attending the patient and team (approximately 60.6\%), followed by referencing the display for anticipatory decisions (25.65\%), and the patient monitor (9.89\%).
In addition, both leaders spent time discussing among themselves in these situations, which may have excluded other team members' input.

During Session 3, the team leaders focused the majority of their time on the patient and team (45.05\%), followed by the patient monitor (40.03\%), and the shared display (13.67\%).

\subsection{Educator Observations}
We reviewed recordings with two clinical educators at the hospital to explore their perspective of teams using Visual TASK.
The main problems noted by educators were recognition issues with the Kinect in Sessions 1 and 2.
The gesture for getting the system to recognize the team leader (raising their hand over their head) imposed issues depending on placement, bed size, and organization of the room.
Participants frequently stepped out of view of the Kinect when attending to other tasks and had trouble getting the Kinect to recognize their hand.
Occlusion factors imposed by the environment and teammates also caused issues with the use of the device.

One such occlusion factor was caused by the patient bed in Sessions 1 and 2.
This bed included a large railing and that created a box around the patient mannequin (see Figure \ref{fig:fig3} A).
The occlusion caused by the bed required the team lead to constrain their movements  to a specific position in the room in order to be recognized by the tool.

In contrast, we did not observe interaction issues in Session 3, which used a touch screen instead of the Kinect.
The shift in focus from interaction caused team leaders to focus much more on other objects in the room, like the patient monitor.
The educators noted that they frequently train clinicians (not necessarily the team leader) to check the patient monitor often for changes in vitals.
Educators speculate that this change in fixation may be either due to clinicians fixating on the next most familiar object in the hopes of it providing new information (the monitor in this case), or that clinicians fixate on objects in general in high-stress situations (not only reference guides).

In addition, educators observed instances where clinicians were unaware of next steps despite having our guide available to them.
For example, a physician acting as team leader correctly identified the heart rhythm name out loud, but was unaware if that particular heart rhythm is shockable.
While the physician eventually proceeded through the code correctly, educators noted this may not be an inherent problem with a lack of any information with cognitive aids, but rather, a potential lack of knowledge of the individual.
Educators speculate that it may be possible that clinicians may default to the algorithms for answers, even if it is not there, and only engage in critical thinking when the algorithm fails.

\subsection{Participant Feedback}

In the two sessions that used the Kinect, participants faced problems with tool usability. 86\% of participants (13/15) thought that Visual TASK with the Kinect distracted the team leader. 66\% of clinicians  in the first two sessions (10/15) proposed changing the interaction modality to a touchscreen device.
Participants criticized the fact that the Kinect distracted the team leader when it was used.

In their written feedback, team leaders noted that gestures made it difficult to respond as quickly as they would have liked.
Three participants noted that the ``\textit{controller did not respond quickly}.''
One clinician noted that the system ``\textit{distracted from [the] leader's role to ensure other aspects of code were  running appropriately - for [example], as the leader, I would notice if CPR was not being adequately performed, but because I was concentrating on the system and getting it to recog[nize] my hand, I didn't see that my team wasn't doing 15:2 compressions.}''

Team leaders also noted that the Kinect restricted their movement to one space, taking their attention and focus away from the patient and team.
Participants noted that due to this, the physical space might be impacted during a code, and need to be refined.
Thus, gesture-based interaction tools like the Kinect may be unfit for training in high-stress simulations.

However, 70\% of all participants (16/23) responded positively to the shared display aspect of Visual TASK.
Participants noted that the tool provided ``\textit{quick access and visualization for all members}'' and that ``\textit{that everyone could be on the same page.}''
Participants liked the ability to verify each others' tasks and know where they were as a team in the algorithm.
One physician noted that the step-by-step presentation of tasks ``\textit{allows for easier visualization of what the next step is, and ensures that certain steps don't get missed.}''
All participants in Session 3 (where the touchscreen was employed) did not view the tool as distracting compared to the first two sessions that used the Kinect.

Individuals in the team leader role also said that the tool helped them to deduce future steps during the code.
One team leader noted that, ``\textit{It helped to determine where to go next in the algorithm},'' while another commented that it ``\textit{showed [them] step-by-step what to do.}''
Other participants mentioned that they found the tool ``\textit{straightforward, [and took] less time finding the info needed.}''
Thus, a design focused on the situation facing clinicians in resuscitation may be better suited for attending to tasks and promote a more cohesive mental model.

\section{Discussion}
Our study yielded a number of insights for the design of collaborative cognitive aids in healthcare situations.
First and foremost, we found that interaction modalities like the Kinect, while potentially helpful for certain areas of acute care like the operating room \cite{Ruppert2012, o2014}, may not yet be ready fast-paced environments like acute care resuscitation.
Even as a training tool, the Kinect is not able to accurately keep up with the range of movement teams need in order to be effective, especially compared to applications similar to those in the operating room \cite{o2014}.
Despite the desire for educators to deploy tools like the Kinect to help train clinicians to keep their heads up, there are simply too many variables that can factor into the placement and tracking for it to be used reliably in acute simulation.

Thus, while the Kinect may be helpful for clinicians in low intensity environments to operate tools sterily, using the Kinect in higher-paced situations like resuscitation could lead to more patient deaths.
Recognition issues with the Kinect, and possibly other such interaction devices, simply cannot keep up with the needs of the team.
As it stands, the Kinect only creates additional situational awareness complications with its use.

However, there may be some environments where the Kinect is an appropriate interface modality.
It may be usable in more static environments, like classrooms.
The clinical educators believed that using the tool earlier in education, like in medical school, might help train clinicians to form better habits in code situations.

In our study, the touch-screen version of Visual Task was perceived to be more user-friendly in a high-paced environment, despite an initial hesitation that it might create a division of attention.
Clinical educators noted that it might be helpful to consider alternate interaction modalities for the team leader moving forward, such as tablets.

Clinicians responded favorably to the redesigned PALS reference aid and shared display aspects of Visual TASK.
Team members liked the ability to visualize information that pertains to each of their tasks, without the clutter of information spread across multiple pages as it is on PALS cards.
In addition, clinicians noted the inclusion of the history chart is beneficial for new entrants to the scene, potentially alleviating distractors or communication challenges that cause preventable medical errors.
Generally, participants viewed the layout of the system as more user-friendly 

We also observed a number of other contributions that may impact the way educators may want to approach training algorithms like PALS.
For example, we observed that during simulated codes, clinicians fixated on items like the patient monitor often, when standard PALS cards were not physically present.
According to educators, the patient monitor, code team, and patient are all legitimate areas of focus and the team as a whole needs to attend to all three simultaneously.  
Educators note that teams appear to work best when a team member focuses on the monitor, other team members focus on the patient, and the team leader focuses largely on the team with frequent (but not fixed) reference to the patient and monitor.  
The fact that the monitor took up a large amount of attentional time of the team leader might indicate that clinicians may fixate on objects that are either familiar, or those that they hope might clue them into situations they are not comfortable with.

However, fixating on objects in the room, rather than the team or the patient, tends to distract or move the team leader's field-of-vision away from the scene.
A tool like Visual TASK might prove useful to help clinicians train to keep peripherals focused on a scene in a low stress environment.

Furthermore, we also observed that when team leaders use reference aids (including Visual TASK), they do so expecting the tool to provide the key and answer to the problem they are facing, and may lack the knowledge or understanding of the condition in front of them.
For example, in Session 3, clinicians did not know if a patient condition (Ventricular Tachycardia) was a shockable condition or not.
Thus, if there is no reference guide available to a team leader in a scenario, the inability to immediately know whether or not a rhythm is shockable undermines the goals of the PALS course and algorithms.
This is a knowledge-based issue with clinicians undergoing training that may need to be addressed earlier on in education, or in PALS/ACLS courses \cite{PALS, ACLS}.

Up until this point, educators could not observe this phenomena because the paper-based PALS reference guide is the only evidence-based guide for clinicians to use in pediatric resuscitation.
By changing the tool clinicians used from the paper guide, clinicians attempted to find solutions to problems that should be a key part of their knowledge.
This underlies a key aspect of resuscitation that clinicians need to understand: the critical relationships between the written algorithms and the patient's condition.

Overall, it appeared that if the team leader expressed a clear understanding of next steps, the rest of the team did not need to reference the shared guide.
However, our mock codes did not emulate some problems that may occur when codes start.
For example, in real resuscitation codes, new people may enter the scene at any given point in time.
Thus, individuals, including the physician who may eventually take on the team leader role when they arrive, may have no context of the tasks performed upon entering the scene.

Both the clinical educators and participants noted that with practice, tools like Visual TASK may be advantageous for training clinicians to keep their peripheral attention available.
They also felt it could be a tool to help clinicians form  and maintain a common mental model of a situation.

The fact that clinicians were able to successfully employ Visual TASK in a high-stress situation points to the advantages of designing cognitive aids that focus adapt to on the situations clinicians face.
In contrast, PALS cards, which provide the entire algorithm across multiple pages, may hinder SA, especially since all information needs to be verbally conveyed.
Tools like Visual TASK may even be able to be incorporated into electronic health record systems in the future, enabling clinicians to enter ``code modes'' to automatically document completed tasks in real-time.

Moving forward and expanding upon this work, we will evaluate the effect of Visual TASK on SA and workload in resuscitation compared to currently used PALS guides, as well as with different interaction modalities.
The goal of these evaluations is to explore whether tools like Visual TASK can improve how teams visualize, share, and coordinate their actions in resuscitation, and later, in other high-paced, high-stress situations.

In addition, we plan to evaluate the tool across the three institutions that aided in the co-design of the tool.
This can help us understand additional insights into collaborative cognitive aid design among a wide-breadth of healthcare environments.
Our hope is our research can aid in improving situational awareness in resuscitation, and later, become extensible to other areas of collaborative healthcare like telemedicine.
This can be advantageous for hospitals which lack experts in resuscitation, allowing others to remotely call in and help guide clinicians during codes.

\section{Limitations}
We acknowledge there are limitations to this work. First, our study represents only a select group of participants from one institution across three simulated resuscitation scenarios. Second, we acknowledge that this work explores our tool with only the Kinect and touchscreen, and that additional studies are required to assess how the tool compares to traditional PALS guides. Finally, we acknowledge that this work is qualitative in nature, and additional studies exploring the concept of fixation are required to know how collaborative cognitive aids affect teams in resuscitation. Despite these limitations, we believe our work is an important contribution to the field of collaborative healthcare technology design due to the fact that it explores notable changes to standardized guides in the field, with the aim of improving outcomes.

\balance{}

%
\bibliographystyle{abbrv}
\bibliography{sigproc}  
%
%


\end{document}